%% file: main.tex
\DocumentMetadata{}
\documentclass[sigconf]{acmart}

\usepackage{graphicx}
\usepackage{subfigure}
\usepackage{multirow}
\usepackage{float}

\AtBeginDocument{%
  \providecommand\BibTeX{{%
    \normalfont B\kern-0.5em{\scshape i\kern-0.25em b}\kern-0.8em\TeX}}}

\copyrightyear{2026}
\acmYear{2026}
\setcopyright{cc}
\setcctype{by-nc-nd}
\acmConference[CHI EA '26]{Extended Abstracts of the 2026 CHI Conference on Human Factors in Computing Systems}{April 13--17, 2026}{Barcelona, Spain}
\acmBooktitle{Extended Abstracts of the 2026 CHI Conference on Human Factors in Computing Systems (CHI EA '26), April 13--17, 2026, Barcelona, Spain}
\acmDOI{10.1145/3772363.3798373}
\acmISBN{979-8-4007-2281-3/2026/04}

\begin{document}

\title{InterPilot: Exploring the Design Space of AI-assisted Job Interview Support for HR Professionals}

\author{Zhengtao Xu}
\email{xuzhengtao@u.nus.edu}
\orcid{0009-0003-4429-4768}
\affiliation{%
  \department{Computer Science}
  \institution{National University of Singapore}
  \city{Singapore}
  \country{Singapore}
}

\author{Zimo Xia}
\email{zimo.xia@u.nus.edu}
\orcid{0009-0007-6308-3439}
\affiliation{%
  \department{Faculty of Science}
  \institution{National University of Singapore}
  \city{Singapore}
  \country{Singapore}
}

\author{Zicheng Zhu}
\email{zicheng@u.nus.edu}
\affiliation{%
  \department{Computer Science}
  \institution{National University of Singapore}
  \city{Singapore}
  \country{Singapore}
}

\author{Nattapat Boonprakong}
\email{nattapat.boonprakong@nus.edu.sg}
\affiliation{%
  \department{Computer Science}
  \institution{National University of Singapore}
  \city{Singapore}
  \country{Singapore}
}

\author{Yu-An Chen}
\email{annchen1@g.harvard.edu}
\affiliation{%
  \institution{Harvard University}
  \city{Cambridge}
  \state{Massachusetts}
  \country{USA}
}

\author{Rabih Zbib}
\email{rabih.zbib@avature.net}
\affiliation{%
  \institution{Avature}
  \city{Barcelona}
  \country{Spain}
}

\author{Casimiro Pio Carrino}
\email{casimiro.carrino@avature.net}
\affiliation{%
  \institution{Avature}
  \city{Barcelona}
  \country{Spain}
}

\author{Yi-Chieh Lee}
\email{yclee@nus.edu.sg}
\orcid{0000-0002-5484-6066}
\affiliation{%
  \department{Computer Science}
  \institution{National University of Singapore}
  \city{Singapore}
  \country{Singapore}
}
\begin{abstract}
  \input{section/0Abstract}
\end{abstract}




\maketitle

\begin{figure*}[t]
  \centering
  \includegraphics[width=1\textwidth]{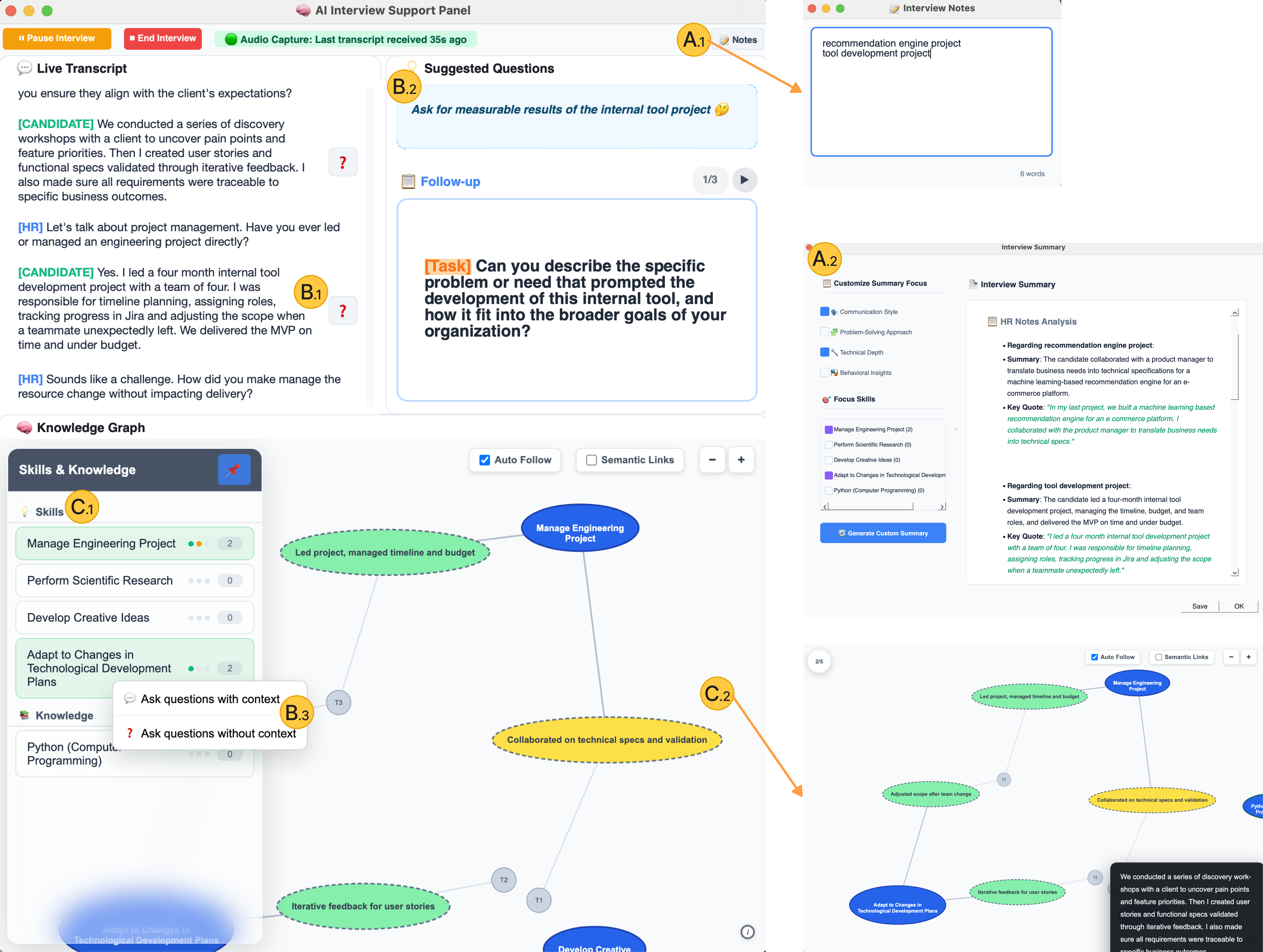}
  \caption{Overview of \textsc{InterPilot} system interface. The system supports HR professionals' interview workflow through three main components: (A) Intelligent Note-taking \& Summary, (B) Adaptive Question Generation, and (C) Real-time Skill \& Evidence Mapping. Sub-labels (e.g., A.1, B.1) correspond to specific features detailed in Section~\ref{sec:user_interface}.}
  \Description{Screenshot of the InterPilot interface. The left panel shows a real-time transcript, the top-right area displays AI-generated follow-up questions, and the bottom panel visualizes a skill--evidence knowledge graph. Insets on the right illustrate the note-taking panel and the post-interview summary view, with callouts (A.1--C.2) highlighting specific features within each component.}
  \label{fig:system}
\end{figure*}

\input{section/1Introduction}

\input{section/2FormativeStudy}

\input{section/3InterPilot}

\input{section/4UserStudy}

\input{section/5Discussion}

\begin{acks}
This research was supported by Avature Group Limited (A-8002873-00-00). We thank all the HR professionals who participated in our studies. We also thank the reviewers for their valuable comments and suggestions that helped improve this paper.
\end{acks}

\newpage
\bibliographystyle{ACM-Reference-Format}
\bibliography{reference}

\appendix
\newpage
\input{section/Appendix}
\end{document}

%% file: section/0Abstract.tex
Recruitment interviews are cognitively demanding interactions in which interviewers must simultaneously listen, evaluate candidates, take notes, and formulate follow-up questions. To better understand these challenges, we conducted a formative study with eight HR professionals, from which we derived key design goals for real-time AI support. Guided by these insights, we developed \textsc{InterPilot}, a prototype system that augments interviews through intelligent note-taking and post-interview summary, adaptive question generation, and real-time skill–evidence mapping. We evaluated the system with another seven HR professionals in mock interviews using a within-subjects design. Results show that \textsc{InterPilot} reduced documentation burden without increasing overall workload, but introduced usability trade-offs related to visual attention and interaction complexity. Qualitative findings further reveal tensions around trust and verification when AI suggests highly specific technical questions. We discuss implications for designing future real-time human–AI collaboration in professional settings, highlighting the need to balance assistance granularity, attentional demands, and human agency.

%% file: section/1Introduction.tex
\section{Introduction}

Recruitment interviews are cognitively demanding interactions for Human Resource (HR) professionals. While structured interviews are widely advocated for their standardization and ease of execution \cite{campion1997review, levashina2014structured}, in practice most interviews remain semi-structured to preserve flexibility and natural conversation \cite{van2002structured, nolan2014need, lievens2004empirical, highhouse2008stubborn}. This flexibility, however, places substantial cognitive demands on interviewers, who must simultaneously listen, take notes, assess competencies, and plan follow-ups. Such multitasking can lead to cognitive overload, reduced attention and increased susceptibility to bias in candidate assessment \cite{appelbaum2008multi, kith2018effect}. It is also challenging for interviewers to formulate probing questions about areas beyond their expertise, often resulting in superficial assessments or missed critical information \cite{rousseau2011becoming}. Furthermore, unlike general semi-structured interviews, recruitment interactions operate under unique constraints: interviewers must maintain strict evaluative consistency across candidates to ensure fairness and objectivity \cite{campion1997review}.
Without adequate support, these inherent human limitations create a key bottleneck for conducting effective and fair interviews.

To address these limitations, the recruitment industry has increasingly integrated AI technologies, yet their use remains largely focused on asynchronous workflows \cite{liu2025interview, liu2025envisioning}. Leading platforms, such as Eightfold.ai\footnote{\url{https://eightfold.ai/}} and Paradox\footnote{\url{https://www.paradox.ai/}}, primarily  automate pre- and post-interview tasks, including candidate matching, resume screening, automated scheduling, and asynchronous assessments. 
In contrast, support for live interviews is minimal. Existing real-time tools, such as Otter.ai\footnote{\url{https://otter.ai/}}, emphasize transcription and note organization rather than domain-specific support. What is missing in the current HR landscape is real-time, context-aware cognitive support to help interviewers understand, ask appropriate questions, and enhance the objectivity of evaluation.

While prior work has explored human-AI collaboration in decision-making, applying real-time AI assistance to high-stakes, socially complex settings like live interviews remains underexplored. To address this gap, we first conducted a formative study with eight HR professionals to identify the specific challenges they face during live interviews. Guided by the resulting design goals, we developed \textsc{InterPilot}, a prototype system that augments human judgment through AI-generated note-taking and post-interview summary, adaptive question generation, and real-time skill-evidence mapping (Figure~\ref{fig:system}). We evaluated the system with another seven HR professionals in mock interview sessions. Our finding revealed a nuanced design tension: while \textsc{InterPilot} helped offload cognitive burden, scaffold questioning, and improve perceived objectivity, it introduced new challenges in visual attention management and trust calibration.

Our work provides three primary contributions to the HCI community: 1) we empirically characterize the challenges HR professionals face during live recruitment interviews; 2) we present a design prototype that illustrates how AI can collaborate with interviewers; 3) we reveal two design tensions in real-time AI support—between functional assistance and attentional cost, and between prescriptive ``scripting'' and diagnostic guidance—offering design implications for real-time AI support in high-stakes decision-making.

%% file: section/2FormativeStudy.tex
\section{Formative Study}

To identify the design space for AI-assisted interviewing, we conducted semi-structured interviews with eight HR professionals (4 female, 4 male; coded as P1–P8), aged 28--51 ($M=37.63,SD=8.46$), with 2--28 years of recruitment experience ($M=11.75, SD=10.12$) across diverse sectors including education, consulting, banking, energy and technology. Interviews focused on challenges in conducting interviews and opportunities for AI support. We analyzed the interview transcripts using thematic analysis \cite{braun2006using}. The results indicated that HR professionals encounter several challenges, especially in managing the high cognitive load from multitasking, formulating deep and adaptive questions in real-time, and maintaining objectivity when evaluating candidates. From these findings, we identified three key design goals (DGs) for our system:

\textbf{DG1: Offload Cognitive Burden from Multitasking.} 
Participants described interviews as mentally taxing due to the tension between maintaining conversational flow and capturing details. As P5 noted, \textit{``I need to try to remember everything... and also assess the candidate at the same time.''} P8 further emphasized that human interviewers \textit{``really cannot''} multitask effectively, leading to mental fatigue. 
Crucially, participants identified repetitive tasks as the primary target for AI intervention. P7 explicitly stated, \textit{``Administrative tasks like taking notes... should be automated. Professionals should concentrate on tasks related to their core expertise.''}
Consequently, a primary goal is to streamline repetitive administrative tasks so HR can focus on active listening. This goal aligns with prior literature demonstrating that interviewer cognitive load negatively affects assessment quality and that structured support tools can improve interviewer performance \cite{appelbaum2008multi, campion1997review}.

\textbf{DG2: Scaffold Context-Aware Question for Deeper Probing.} 
Another major challenge identified was the difficulty in generating deep, adaptive questions in real-time, due to time constraints (P4) and limited domain knowledge (P1). P3 admitted to giving \textit{``generic feedback''} because they struggled to understand the nuances of a candidate's story. 
Participants expressed a desire for AI to bridge this gap. P5 envisioned an interaction where \textit{``we can tell AI that we want to assess the autonomy level... and it can help us generate deeper questions accordingly.''} Similarly, for technical roles, P6 noted that AI support could help them \textit{``roughly ask technical questions''} to make the interaction smoother.
Therefore, the system should empower probing by acting as a ``co-pilot'' that adaptively suggests follow-up questions. This draws on principles of mixed-initiative human-AI collaboration \cite{horvitz1999principles, inkpen2023advancing}, using AI to bridge domain expertise gaps while keeping the human interviewer in control of the conversation.

\textbf{DG3: Objectify Skill Assessment with Structural Evidence.} 
Evaluating candidates' skills is highly subjective for different HRs. P1 highlighted that \textit{``we are humans, so many things depend heavily on individual judgment.''}
Furthermore, high workloads can trigger \textit{``confirmation bias,''} where recruiters subconsciously look for specific evidence to support their initial impressions (P6).
To address this, the system should facilitate evidence-based evaluation by enabling real-time tagging of candidate responses against specific criteria. By shifting assessment from subjective ``gut feelings'' to a structured framework, this goal implements established interview techniques \cite{campion1997review, levashina2014structured} and acts as a cognitive forcing function \cite{buccinca2021trust} against confirmation bias.

%% file: section/3InterPilot.tex
\section{InterPilot System} 
\label{sec:user_interface}

To address our design goals, we designed and developed a prototype system, \textsc{InterPilot}, to support HR professionals during interviews.
As illustrated in Figure \ref{fig:system}, the system integrates three components (A, B, C) as follows:

\textbf{A. Intelligent Note-taking and Post-interview Summary.} InterPilot automates the synthesis of interview data. During the interview, the \textit{Note-taking Widget} (\textbf{A.1}) allows HR professionals to simply jot down high-level keywords instead of verbatim logging. 
Upon completion, the system synthesizes these keywords with the full transcript and skill analysis to generate a structured \textit{Summary Report} (\textbf{A.2}). This component allows interviewers to streamline documentation (\textbf{DG1}) and provides a standardized basis for final candidate assessment (\textbf{DG3}).

\textbf{B. Adaptive Question Generation.} 
InterPilot offers question generation in three modes: (\textbf{B.1} \textit{Deep Probing}) users can click the ``?'' button to generate follow-up questions based on the STAR framework (Situation, Task, Action, Result) \cite{byham1989targeted}; (\textbf{B.2} \textit{Contextual Probing}) users can obtain high-level directions from the blue area for subsequent interview questions from AI; and (\textbf{B.3} \textit{Targeted Probing}) users can select a skill from the list to generate a question that assesses it. 
This component helps scaffold context-aware questions (\textbf{DG2}) and reduces the cognitive burden of formulating inquiries on the fly (\textbf{DG1}).

\textbf{C. Real-time Skill and Evidence Mapping.} 
The \textit{Skills List} (\textbf{C.1}) serves as a progress tracker, helping HR professionals identify which areas need further probing during the interview.
Complementing this, the \textit{Knowledge Graph} (\textbf{C.2}) dynamically visualizes the connections between the required skills (\textbf{Blue Nodes}), extracted evidence (\textbf{Dashed Nodes}, color-coded to denote relevance strength to the skill: \textbf{Green} for high, \textbf{Yellow} for medium), and specific transcript segments of the conversation (\textbf{Grey Nodes}). This skill-evidence-transcript mapping helps reduce cognitive load (\textbf{DG1}) and support objective assessment (\textbf{DG3}).

%% file: section/4UserStudy.tex
\section{User Evaluation Study}


\subsection{Study Design and Measurements}
For the evaluation, we invited a new group of seven HR professionals (6 female, 1 male; coded as H1–H7), aged 35--50 ($M=40.29, SD=5.18$), with 6--20 years of recruitment experience ($M =13.14, SD = 5.78$). Their backgrounds spanned sectors including technology, consumer goods, manufacturing, and education.
Our study received ethics approval from the university's IRB.

We employed a within-subjects design where each participant conducted two short mock video interviews (10–15 minutes each) for a \textit{Software Developer} position through Zoom Meeting\footnote{\url{https://www.zoom.com/}}. One session used \textsc{InterPilot}, while the other used a \textit{Baseline System} (live transcript only). Two research assistants acted as confederates playing the candidate role. They followed a predefined candidate profile and were trained to provide relatively consistent responses across interview sessions, a common practice in HCI studies to control interaction variability while preserving natural dialogue \cite{shi2019accessible, convertino2004laboratory, hyde2014conversing}.

Prior to the study, participants received an informed consent form, a tutorial video, and the job description. At the start of the session, they were given a hands-on tutorial to familiarize themselves with the assigned system's features. To mitigate learning and order effects, the presentation order of the two systems was counterbalanced across participants. After completing the first interview, participants filled out a survey and then proceeded to the second condition. The study concluded with a semi-structured interview to gather in-depth feedback on their experience and the system's impact on their workflow. The entire study took about 60 minutes.

We collected both quantitative and qualitative measures. After each interview, participants completed: (1) the \textbf{System Usability Scale (SUS)} \cite{brooke1996sus} to assess usability and (2) the \textbf{NASA-task load index (NASA-TLX)} \cite{hart1988development, hart2006nasa} to measure workload. Qualitative data were derived from the post-study semi-structured interviews, which were also analyzed using thematic analysis \cite{braun2006using}, focusing on the system's utility, interference, and design implications.

\subsection{Results}


\subsubsection{Mental Workload and Usability Trade-offs}
\label{sec:workload}

InterPilot did not impose a higher cognitive burden compared to the simple baseline. A paired-samples t-test on NASA-TLX overall workload scores showed no statistically significant difference between two systems 
(Baseline: $M=3.62, SD=1.07$ vs. \textsc{InterPilot}: $M=3.52, SD=1.09$), 
$t(6)=0.17, p=0.87, d=0.07$. This implies that InterPilot's features did not impose significant mental demand compared to a transcript-only baseline. 
Qualitative feedback echoed this finding. Participants praised the automated summary for reducing the ``scribbling burden''. As H7 noted, \textit{``I don't need to write down everything... I can focus on listening and maintaining eye contact.''} H6 added that the system serves as a reliable backup that \textit{``reduces memory load.''}

In contrast, usability scores revealed a clear trade-off. A paired-samples t-test on SUS scores showed a significant difference between two systems (Baseline: $M=82.1, SD=6.2$ vs. \textsc{InterPilot}: $M=72.5, SD=10.1$), $t(6)=2.65, p=0.04, d=1.00$. InterPilot's lower SUS mean score reflects the increased interaction and visual complexity introduced by real-time AI features, which participants found demanding to manage alongside ongoing conversation. As H7 commented, \textit{``having all the different options was a bit overwhelming.''} H6 further elaborated on the visual demand: \textit{``A lot of things are happening... transcript is populated, graph is auto-followed... It will be a lot for someone to deal with.''} 

\subsubsection{Scaffolding Question Depth \& Breadth}
Qualitative data suggested that participants valued InterPilot as a ``safety net'' that scaffolded their questioning strategy, particularly for navigating ``stuck moments''. H3 described the \textit{Adaptive Question Generation} component as a true co-pilot: \textit{``It gives me ideas when I’m stuck... serves as a prompt for me to dig deeper.''} Beyond general depth, the system played a crucial role in bridging domain knowledge gaps. H1 specifically highlighted its utility for probing questions to assess areas beyond the interviewer's expertise: \textit{``This tool will be useful for hiring managers that are not so well trained... if you ask me to probe on Python skills then I may not be able to pass.''}

Participants, however, expressed concerns that AI-suggested questions were highly specific and dictative. 
H7 showed hesitation when AI suggestions became too technical to verify: \textit{``Sometimes... they were very specific... I don’t even know if that question makes sense, so I don’t want to ask it.''} 
In response, some participants suggested a shift from direct ``scripting'' to ``diagnostic'' support. H2 proposed that instead of generating full questions, the system could simply flag areas of concern: \textit{``maybe there can be some suggested points that can help me identify the not-so-good part... be more specific.''} 

\subsubsection{Enhancing Objectivity with Evidence}
The Skill-evidence mapping component proved effective in shifting evaluations away from subjective impressions towards objective facts. Participants reported that the structured framework increased their consistency and sense of agency. 
H3 described the \textit{Skills List} as a \textit{``good reminder''} to ensure she assessed all \textit{``five things that I need to assess.''} H2 highlighted that having ``predefined skills'' clarified her focus: \textit{``I know what I need to assess... that makes me more in control.''} This structure directly helped supported more objective evaluation. As H5 explicitly stated, \textit{``It helps remove bias... because the evaluation is based on facts recorded, not just my feeling.''} 

At the same time, participants expressed more nuanced views toward the \textit{Knowledge Graph} component. While some found the visual summary helpful after the interview, its value during live interaction was questioned. H7 noted that \textit{``the map itself is really helpful for visualizing the person in the summary, but I don’t think it adds that much to the conversation.''} H4 emphasized that the graph was \textit{``better for post-evaluation''} and \textit{``too noisy in real time.''} These mixed responses highlight a key design tension, consistent with the usability trade-offs reported in Section~\ref{sec:workload}.

%% file: section/5Discussion.tex
\section{Discussion}

\subsection{Balancing AI Support and Attention Demand}
Our findings reveal a trade-off between functional richness and attention demands in real-time AI support. While InterPilot did not increase mental workload and was praised for helping to offload the memory burden, its rich real-time visual feedback introduced a noticeable visual attention cost that negatively impacted usability.
This suggests that even when AI support is functionally beneficial, its mode of presentation can compete with users' limited attentional resources during real-time interpersonal tasks.
This tension echoes a long-standing issue in HCI regarding attention management~\cite{linger2024supporting}. \citeauthor{weiser1997coming}'s principle of Calm Technology \cite{weiser1997coming} argues that effective systems should move fluidly between the center and periphery of users' attention, foregrounding information only when necessary. Similarly, prior work has shown that systems demanding frequent visual or cognitive engagement can disrupt primary tasks rather than support them \cite{oviatt2006quiet, adamczyk2004if, iqbal2008effects}. 
Our results suggest that current real-time AI interfaces risk becoming attention-demanding rather than calm, particularly in conversational settings where maintaining social presence is critical.

Importantly, these findings raise a broader design question rather than pointing to a single solution: to what extent should AI systems demand users’ attention in real-time decision-support? While adaptive and mixed-initiative approaches have been proposed as ways to regulate system intervention \cite{horvitz1999principles, jameson2001user}, assistance may still prove intrusive if it requires sustained visual monitoring or manual operation during live interaction.
This tension positions the balance between AI support and attention cost as a central concern in the design of real-time AI systems—not only in terms of what information is provided, but also how and when it enters users’ attentional field. 
Rather than merely prioritizing assistance, future designs of AI interview-support should adapt according to users' attention bandwidth \cite{dingler2017building}, providing support according to users' current cognitive capacities. 

\subsection{Bridging the Verification Gap: From Scripting to Diagnosis}
We identified a critical tension in real-time AI support for interviewing: although InterPilot helped close domain knowledge gaps, highly specific question suggestions could reduce users' willingness to rely on the system. We found that users hesitated to adopt AI suggestions they could not verify. This phenomenon reflects a challenge in human-AI interaction: when systems operate as black boxes, users struggle to calibrate their trust and often reject AI's suggestions \cite{lee2004trust, hoffman2018metrics, dietvorst2015algorithm}. In high-stakes contexts, where errors carry reputational and interpersonal consequences, such under-reliance may be further amplified \cite{frank2024navigating}.

Beyond transparency, our findings point to a second related factor: interviewers' domain expertise shapes how they engage with AI assistance in the moment. Prior work has shown that people rely differently on algorithmic advice depending on their knowledge and ability to evaluate it \cite{inkpen2023advancing}. In our study, interviewers were particularly hesitant when the system proposed technical questions in areas they were unfamiliar with. 
These observations suggest that technical-heavy assistance is not always desirable in real-time recruitment interviews. 
Participants instead favored shifting the AI's role from scripting questions to giving diagnostic support (i.e., highlighting areas of concern or suggesting broad directions for probing). This approach allowed interviewers to formulate follow-up questions based on their own knowledge and style, rather than relying on unfamiliar technical advice.

Together, these results highlight a challenge in real-time decision-making: the tension between domain-bridging assistance and users' ability to verify suggestions in the moment. 
Addressing this gap may require not only explainability but also careful calibration of AI support granularity—preserving human agency, respecting users' expertise boundaries, and ensuring humans remain in control during live interactions \cite{amershi2019guidelines, buccinca2021trust}.

\subsection{Limitations and Future Work}
Our study has several limitations that shape directions for future research. 
First, recruiting HR professionals for in-depth studies proved challenging, and as a result, our current sample was relatively small ($N=7$) and skewed toward senior interviewers. 
Future work will involve a larger, more diverse cohort, especially junior practitioners, to better understand how professional experience affects reliance on and interaction with real-time AI support.
Second, our evaluation was conducted in a controlled mock-interview setting in which trained confederates played the role of candidates rather than real applicants. 
Moreover, the interview scenario focused on a general screening stage, whereas real recruitment processes often involve multiple stages (e.g., technical interviews, managerial rounds, behavioral assessments), each of which may impose distinct cognitive demands and interactional needs. 
Field studies situated in authentic organizational workflows are therefore necessary to examine how InterPilot performs across different interview types and levels of structure.
Third, our study captured short-term interactions and immediate impressions rather than longer-term effects. 
Future work should include longitudinal deployments to investigate how sustained use of real-time AI support influences interviewers’ practices, hiring decisions, and trust calibration over time, as well as broader human–AI dynamics in organizational settings.

%% file: section/Appendix.tex
\section{Statistical Summary}

\begin{table}[h]
\centering

\renewcommand{\arraystretch}{1.2}
\setlength{\tabcolsep}{4pt}

\caption{NASA-TLX workload measures and SUS scores comparing baseline and InterPilot. 
Values are reported as mean (standard deviation). Statistical results are from paired-samples t-tests.}
\label{tab:nasa_sus}
\resizebox{\columnwidth}{!}{
\begin{tabular}{l cc c c c}
\toprule

\multirow{2}{*}{\textbf{Measure}}
& \textbf{Baseline}
& \textbf{InterPilot}
& \multirow{2}{*}{\textbf{\(t(6)\)}}
& \multirow{2}{*}{\textbf{\(p\)}}
& \multirow{2}{*}{\textbf{Cohen's \(d\)}} \\

& \(M\) (\(SD\))
& \(M\) (\(SD\))
& 
& 
& \\

\midrule

Mental Demand      & 4.71 (1.25)& 4.28 (1.38)& 0.51 & 0.63 & 0.19 \\
Physical Demand    & 1.71 (1.89)& 1.71 (1.25)& 0 & 1.00 & 0 \\
Temporal Demand    & 4.43 (1.90)& 4.43 (1.90)& 0 & 1.00 & 0 \\
Performance        & 4.14 (1.46)& 4.71 (1.80)& -1.19 & 0.28 & -0.45 \\
Effort             & 4.14 (1.68)& 3.57 (0.98)& 0.80 & 0.46 & 0.30 \\
Frustration        & 2.57 (1.13)& 2.43 (1.99)& 0.17 & 0.87 & 0.06 \\
\midrule
Overall NASA-TLX   & 3.62 (1.07)& 3.52 (1.09)& 0.17 & 0.87 & 0.07 \\
SUS Score          & 82.1 (6.2)& 72.5 (10.1)& 2.65 & $0.04^{*}$ & 1.00 \\

\bottomrule
\end{tabular}
}
\end{table}

\section{System Implementation}

\textsc{InterPilot} is built as a desktop application using PyQt5 for the GUI, with an embedded QWebEngineView component rendering an interactive D3.js-based knowledge graph for real-time skill-evidence visualization. The backend is powered by a FastAPI server that orchestrates real-time communication between audio capture modules and the GUI via WebSockets, enabling live transcript updates, skill evaluation results, and suggested questions to be pushed to the interface with minimal latency.

For speech-to-text transcription, we utilize AssemblyAI's streaming API to process dual audio streams—capturing HR audio via microphone and candidate audio through the virtual audio device for system audio routing. The AI reasoning pipeline is implemented using GPT-4o as the underlying model, enabling modular agents for skill evaluation, STAR-framework question generation, and interview summarization. Skill-evidence relationships are extracted in structured JSON format and rendered as an interactive knowledge graph in real-time, allowing HR professionals to track demonstrated competencies throughout the interview.

\section{Use case screenshot}

\begin{figure}[b]
    \centering
    \includegraphics[width=1\linewidth]{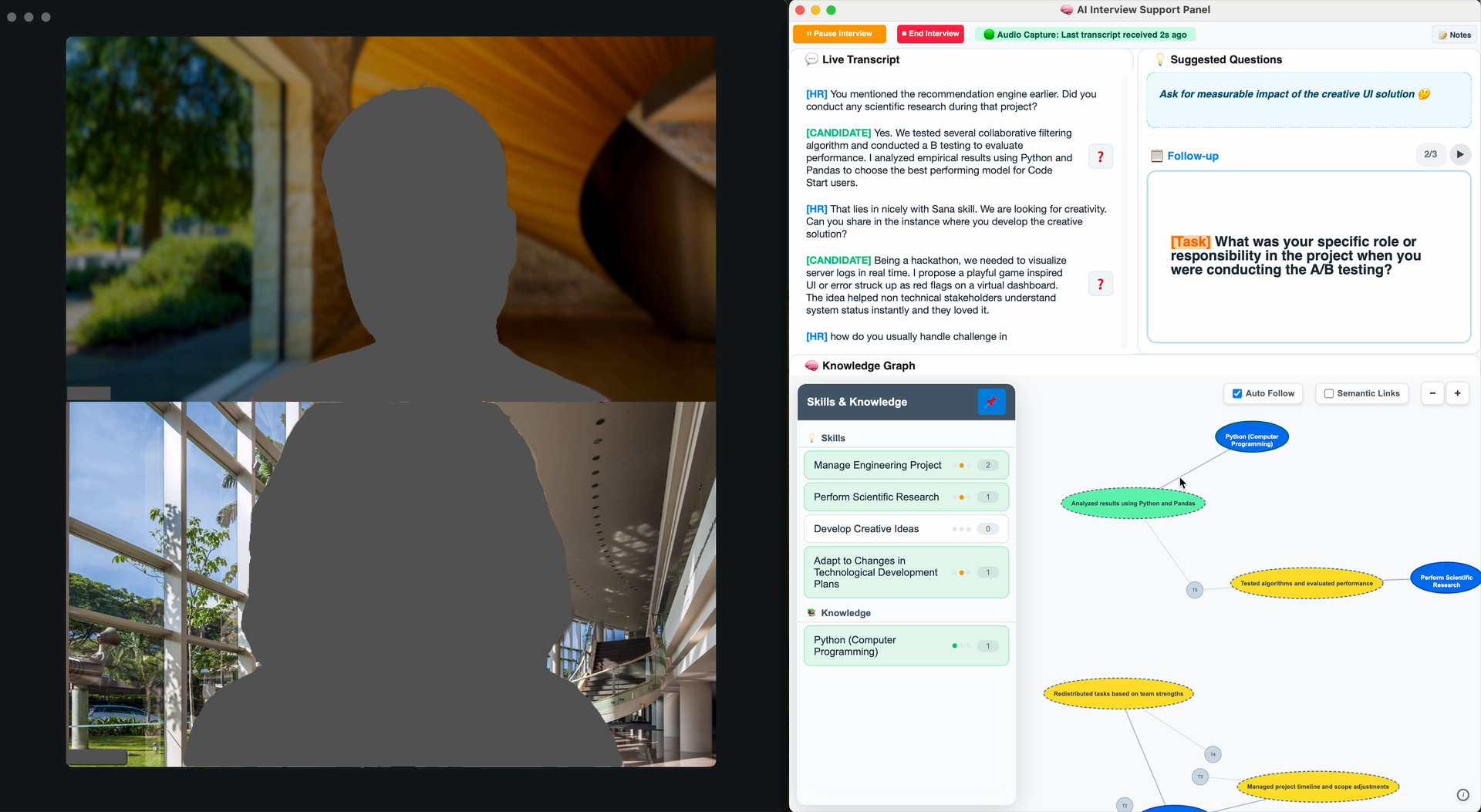}
    \caption{Example interviewer view during a mock interview session. The left panel shows the video conferencing platform (e.g., Zoom), while the right panel displays the InterPilot interface.}
    \Description{Screenshot of a mock interview session showing a split-screen interface. The left side displays a video conferencing platform with the interviewer and candidate video feeds stacked vertically. The right side shows the InterPilot interface, which includes three main components: a live transcript panel at the top displaying the ongoing conversation between the interviewer and candidate; a suggested questions panel providing AI-generated follow-up questions aligned with the STAR framework; and a skills and knowledge graph at the bottom visualizing extracted skills, evidence, and their relationships as interconnected nodes. A sidebar lists identified skills and knowledge areas with indicators of supporting evidence.}
    \label{fig:screenshot}
\end{figure}